\documentclass[english,preprint]{revtex4}
\usepackage[T1]{fontenc}
\usepackage[latin9]{inputenc}
\usepackage{verbatim}
\usepackage{graphicx}
\usepackage{esint}

\makeatletter

\@ifundefined{textcolor}{}
{%
 \definecolor{BLACK}{gray}{0}
 \definecolor{WHITE}{gray}{1}
 \definecolor{RED}{rgb}{1,0,0}
 \definecolor{GREEN}{rgb}{0,1,0}
 \definecolor{BLUE}{rgb}{0,0,1}
 \definecolor{CYAN}{cmyk}{1,0,0,0}
 \definecolor{MAGENTA}{cmyk}{0,1,0,0}
 \definecolor{YELLOW}{cmyk}{0,0,1,0}
 }

\makeatother

\usepackage{babel}

\begin{document}

\title{Coherent dynamics of macroscopic electronic order through a symmetry-breaking
transition.}

\author{R.Yusupov,$^{1}$ T.Mertelj,$^{1}$ V.V.Kabanov,$^{1}$ S.Brazovskii,$^{3}$,
P.Kusar$^{1}$, J.-H.Chu,$^{2}$ I. R. Fisher,$^{2}$ and D. Mihailovic,$^{1}$ }

\affiliation{$^{1}$Dept. of Complex Matter, Jozef Stefan Institute, Jamova 39,
Ljubljana, SI-1000, Ljubljana, Slovenia}

\affiliation{$^{2}$Geballe Laboratory for Advanced Materials and Department of
Applied Physics, Stanford University, California 94305, USA}

\affiliation{$^{3}$ LPTMS-CNRS, UMR8626, Univ. Paris-Sud, Bat. 100, Orsay, F-91405
France}

\maketitle
\textbf{The temporal evolution of systems undergoing symmetry breaking
phase transitions (SBTs) is of great fundamental interest not only
in condensed matter physics, but extends from cosmology to brain function
and finance \cite{topology,Kibble,Eltsov,Finance}. However, the study
of such transitions is often hindered by the fact that they are difficult
to repeat, or they occur very rapidly. Here we report for the first
time on a high-time-resolution study of the evolution of both bosonic
and fermionic excitations through a second order electronic charge-ordering
SBT in a condensed matter system. Using a new three-pulse femtosecond
spectroscopy technique, we periodically quench our model system into
the high-symmetry state, detecting hitherto unrecorded coherent }\textbf{\emph{aperiodic}}\textbf{
undulations of the order parameter (OP), critical slowing down of
the collective mode, and evolution of the particle-hole gap appearing
through the Peierls-BCS mechanism as the system evolves through the
transition. Numerical modeling based on Ginzburg-Landau theory is
used to reproduce the observations without free parameters. The close
analogy with other \textquotedbl{}Higgs potentials\textquotedbl{}
in particle physics\cite{Higgs} gives new insight into hitherto unexplored
dynamics of both single particle and collective excitations through
a SBT. Of particular interest is the observation of spectro-temporal
distortions caused by disturbances of the field arising from spontaneous
annihilation of topological defects, similar to those discussed by
the Kibble-Zurek cosmological model\cite{Kibble}.}

The behaviour of SBTs in many diverse systems is commonly studied
under near-equilibrium (near-ergodic) conditions, where excitations
on all timescales contribute to the process, and the behaviour of
physical quantities through the SBT is described by power laws and
critical exponents. However, when the ordering proceeds non-ergodically,
the situation is fundamentally different. The quasiparticles and collective
boson excitations perceive the homogeneous crystal background as an
effective vacuum. Consequently the ordering takes place as a well-defined
sequence of events in time. For studying the temporal evolution of
elementary excitations through a SBT in condensed matter systems undergoing
second order SBTs, crystals with electronically-driven instabilities
are perhaps the most suitable. Chosing a prominent example amongst
such systems, rare-earth tri-tellurides have an electronic instability
caused by a Fermi surface nesting, leading to a second-order transition
to a broken symmetry charge-density-wave (CDW) ordered state at low
temperature \cite{DIMASI:1994p2472,Ru:2006p2467,STM,Laverock:2005p2473}.
The CDW state is characterized by spatial modulations $\sim\cos(Qx+\phi)$
of the electronic density and of the lattice displacements which give
rise to a complex OP $\Psi\sim\Delta e^{i\phi}$. The elementary bosonic
collective excitations of $\Delta$ and $\phi$ are the amplitude
and the phase modes (AM and PhM), and $\Psi$ may be viewed as a \textquotedbl{}Higgs
field\textquotedbl{},\cite{topology} opening a gap $2|\Delta|$ in
the fermionic spectra. Quasiparticle (QP) and AM excitations in TbTe$_{3}$
have been recently systematically characterized \cite{Yusupov1,ARPES}.
In this paper we focus on TbTe$_{3}$ ($T_{c}=336$ K), but experiments
on a number of microscopically diverse systems (DyTe$_{3}$, 2H-TaSe$_{2}$
and K$_{0.3}$MoO$_{3}$) are also presented demonstrating the range
of behavior observed within this universality class of SBTs.

The main idea realised here is to repetitively quench the system into
the high symmetry state using a short (50 fs) intense \textquotedbl{}destruction\textquotedbl{}
($D$) laser pulse and then monitor the time evolution of reflectivity
oscillations using a pump-probe (P-p) sequence as it freely evolves
through the transition (Fig.1a). (The experimental details, including
data processing are given in the supplementary information (SI)).
The $D$ pulse excites electrons and holes, thus suppressing the electronic
susceptibility at $2k_{F}$ whose divergence is the cause for the
CDW formation. Any asymmetry in the band structure also leads to an
imbalance of the \emph{e} and \emph{h} populations, shifting the chemical
potential and causing a disturbance $\delta\vec{q}$ of the Fermi
surface $\vec{k}_{F}$ according to $n_{e}-n_{h}\propto\mid\delta\vec{q}\mid/\pi$
and destroying the CDW. Following the quench, after initial rapid
quasiparticle (QP) relaxation, we can expect the appearance of topologically
nontrivial local configurations - domain walls, solitons. etc. which
are allowed by the ground state degeneracy with respect to $\phi$
\cite{solitons}. Our technique allows us not only to control and
monitor the emergence of elementary excitations and evolution of $\Psi$
with high time-resolution, but permits a unique experimental detection
of spatio-temporal field distortions arising from domain wall annihilation
events.

\begin{figure}[h]
\begin{centering}
\includegraphics[width=9cm]{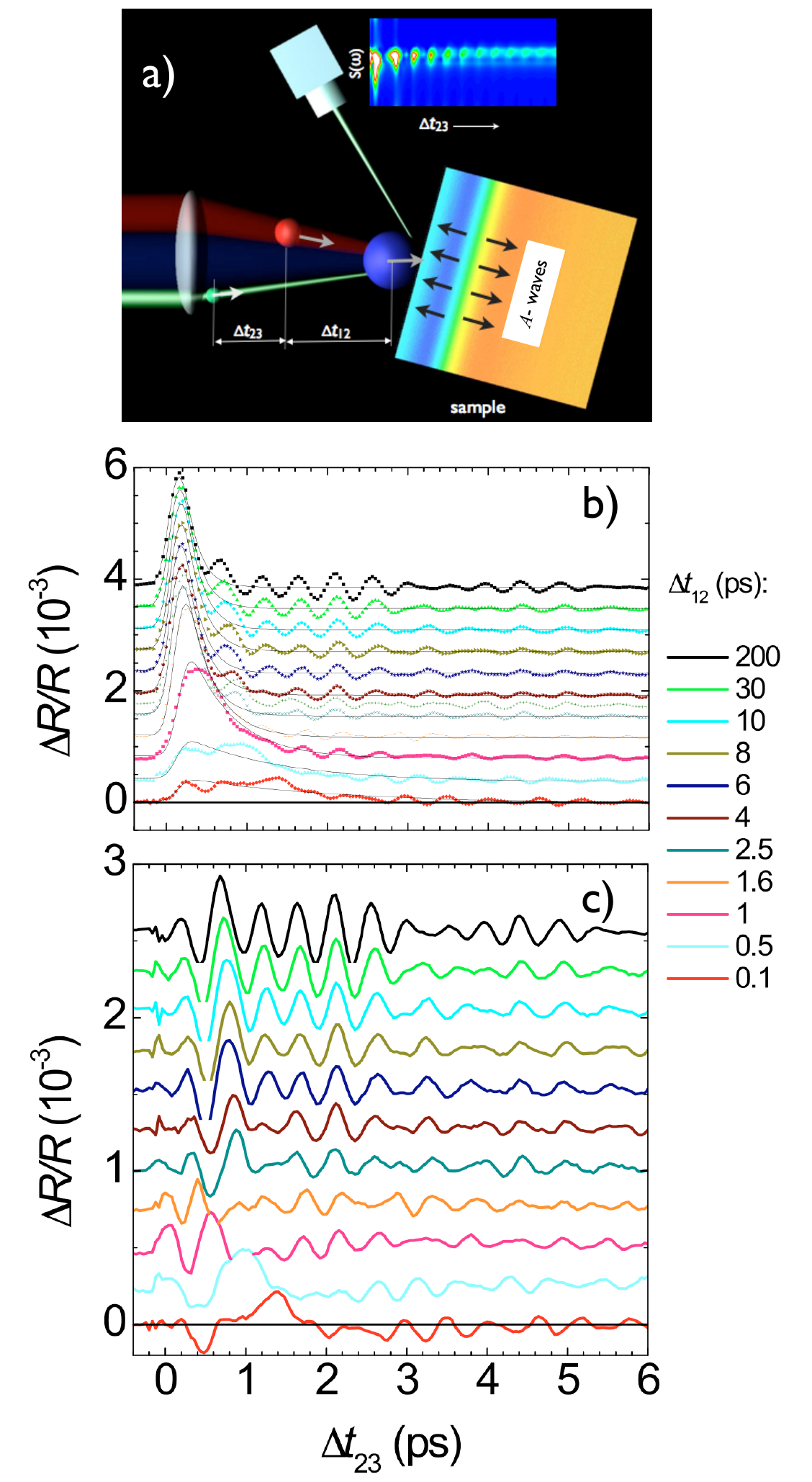} 
\par\end{centering}

\caption{a) A schematic diagram of the timing of the laser pulses: a \textit{destruction}
($D$) pulse (represented by the blue ball) quenches the system, while
a \textit{pump-probe} (\emph{P-p}) sequence probes the reflectivity
at a later time $\Delta t_{12}$. $P$ and $p$ pulses are represented
by red and green balls respectively. b) Raw transient reflectivity
data $\-\Delta R/R$ for different delays $\triangle t_{12}$ (displaced
vertically), showing a QP peak at short times, and OP and coherent
phonon oscillations at longer times c) $\Delta R/R$ with the QP response
subtracted. }

\end{figure}

\begin{figure}
\includegraphics[width=9cm]{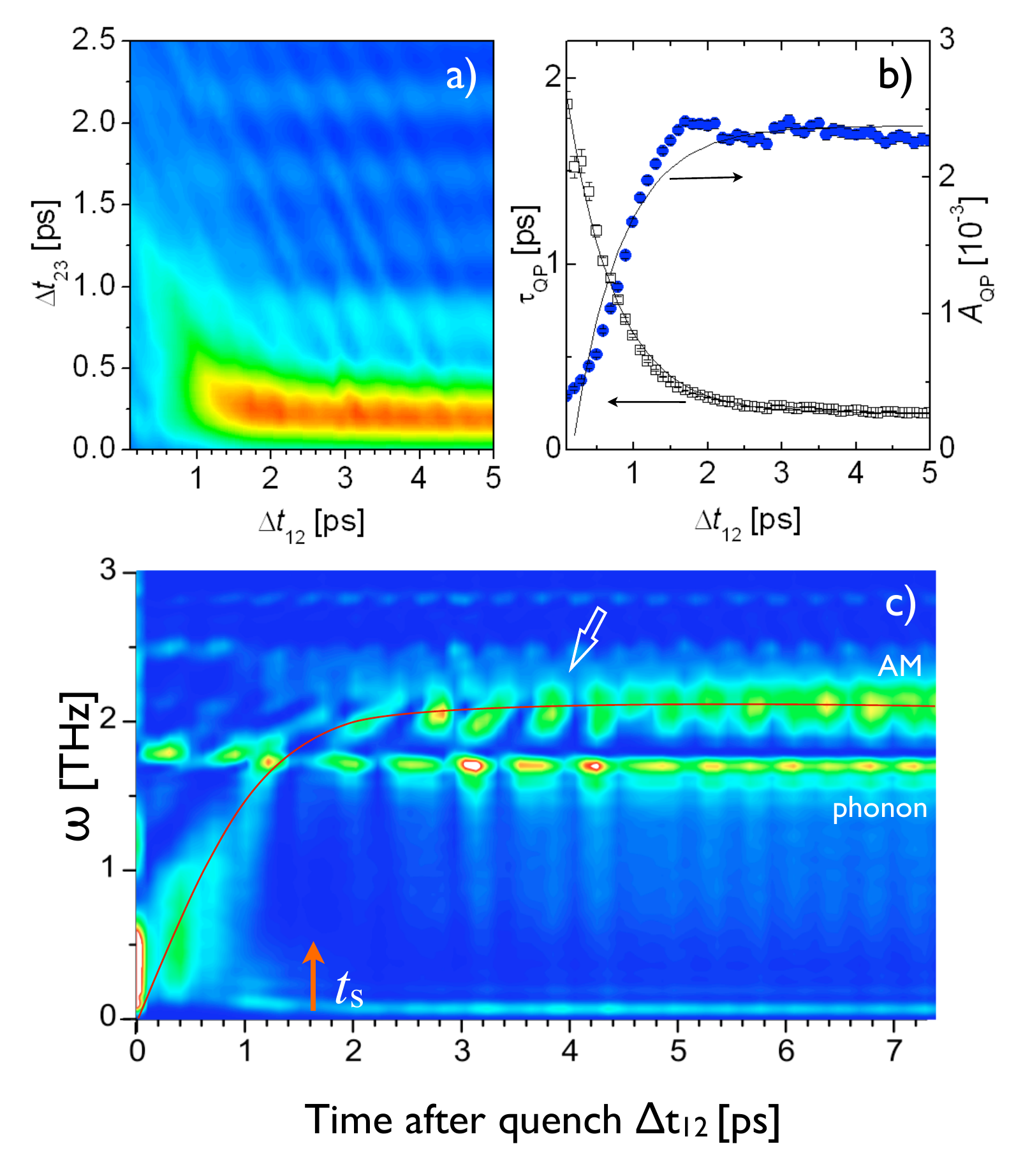}

\caption{$|\Delta R/R|_{QP}$ after the quench as a function of $\Delta t_{12}$,
%%%
(Note the ripples arising from coherent oscillations of the OP.) b)
The QP lifetime $\tau_{QP}$ and the amplitude of the QP response
$A_{sp}$ as a function of $\Delta t_{12}$. A \textit{\emph{single
exponential}} fit to both data sets (shown by the lines) gives $\tau_{\tau_{QP}}=\tau_{A_{QP}}=650\pm50$fs.
c) The FFT power spectra of the data in Fig. 1 c) as a function of
$\Delta t_{12}$ recorded at 100 fs intervals. Note the non-periodic
fluctuations of intensity at around the transition (1.5 ps) and the
strongly asymmetric space-time lineshapes for $\Delta t_{12}=2\sim4$
ps (white arrow). The orange arrow indicates the critical time of
the SBT. The red line is a superimposed plot of a fit to the QP decay
from b) above.}

\end{figure}
Focusing on TbTe$_{3}$, the evolution of the system after a quench
is measured by the transient reflectivity $\Delta R/R$ as a function
of $D-P$ time delay $\Delta t_{12}$ (Fig. 1b)). We can distinctly
see an exponential QP transient at short times $\Delta t_{23}<1$
ps, and an oscillatory response due to AM and coherent phonon oscillations
\cite{Yusupov1} evolving through the SBT. A 2D plot highlighting
the QP response is shown in Fig. 2a). We see that immediately after
the quench the QP peak amplitude $A_{QP}$ is completely suppressed,
indicating the disappearance of a gap. As the QP peak starts to recover,
initially the QP lifetime $\tau_{QP}$ is quite long (a few ps), but
recovers quickly with time. Both $A_{QP}$ and $\tau_{QP}$ recover
to their equilibrium values within 1-2 ps. A single exponential fit
to both $\tau_{QP}$ and $A_{QP}$ is shown in Fig 2b), giving the
QP gap recovery time $\tau_{\tau_{QP}}=\tau_{A_{QP}}=650\pm50$fs.
This is consistent with the previously reported relation $\tau_{QP}\sim1/\Delta(T)$
\cite{KMO,Yusupov1}.

\begin{figure}[h]
\begin{centering}
\includegraphics[width=10cm]{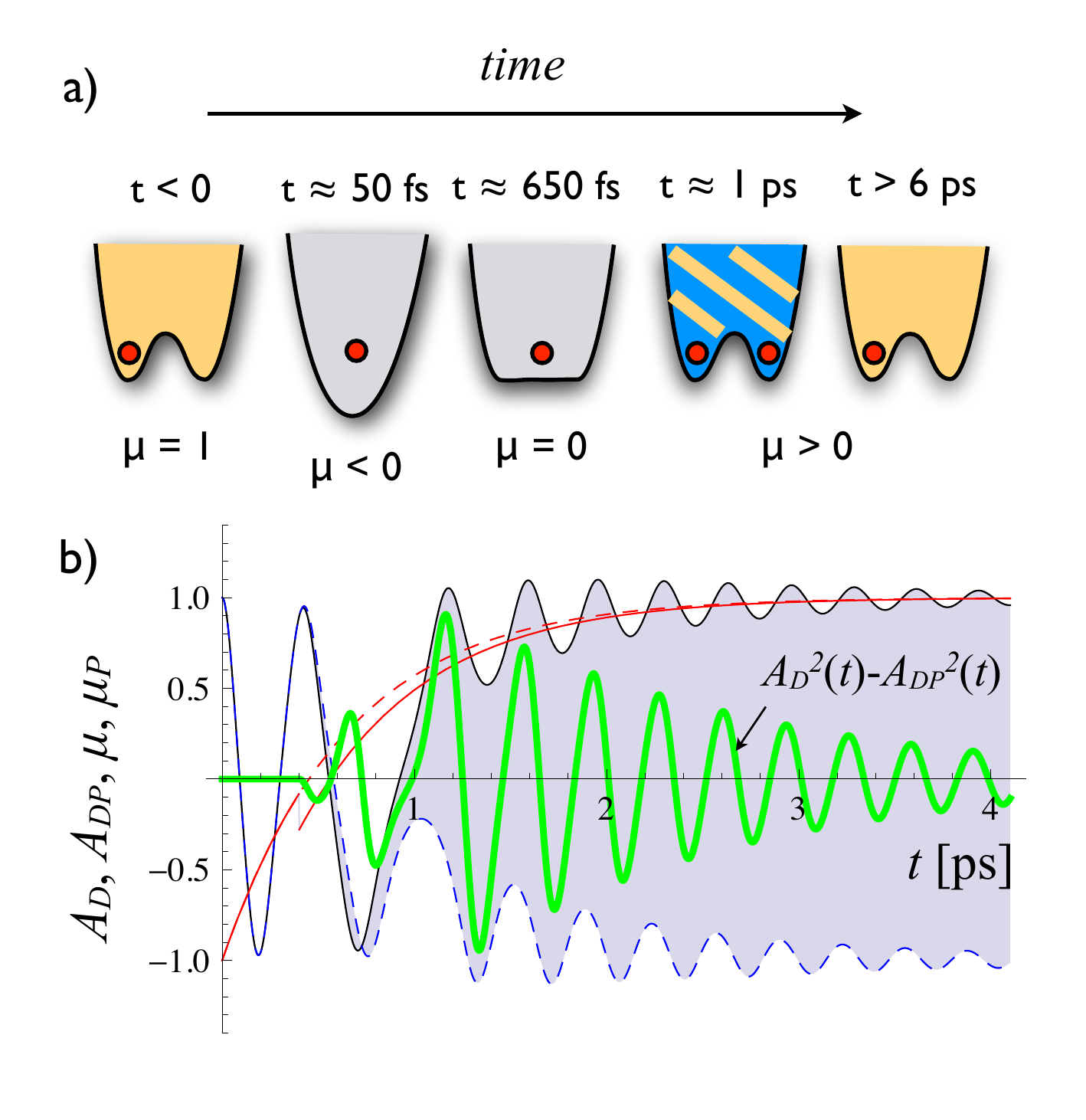} 
\par\end{centering}

\caption{a) The evolution of the potential $U$ (Eq.1) as a function of time.
The system is in the high symmetry state (grey) at very short times
after the quench. The red dot signifies the state of the system, while
the blue/orange potential signifies a topologically mixed broken-symmetry
state. b) The time-dependence of the control parameters $\mu(t)=1-\eta(t)$
(solid red) and $\mu_{P}$$(t)$ (dashed red) calculated with the
experimental value of $\tau_{A_{sp}}=0.65\pm0.$05 ps for$\triangle t_{12}$=0.4
ps. The predicted oscillations of $A(t)$ \emph{with} and \emph{without}
the \emph{P} pulse are shown by the dashed and solid oscillatory blue
curves respectively. The predicted optical response $\Delta R(t)$
is shown by the green curve.  }

\end{figure}

Fig. 1c) shows $\Delta R/R$ for different $\Delta t_{12}$ with the
QP signal subtracted, showing an unusual oscillatory response through
the SBT. The Fast Fourier transform (FFT) power spectra of these data
as a function of $\Delta t_{12}$ are plotted in Fig. 2c). The most
obvious non-trivial observation in the $\omega-\Delta t_{12}$ plots
is that the intensity of the AM fluctuates strongly up to $\sim7$
ps. It then gradually saturates with increasing $\Delta t_{12}$ (not
shown). The fluctuations are irregular at first, showing a distinct
slowing down in the critical region $\Delta t_{12}=1.5$ ps. The AM,
whose frequency in the equilibrium broken symmetry state is 2.18 THz,
shows a dramatic softening for $\Delta t_{12}<2$ ps. In the course
of the system recovery, the AM crosses the 1.75 THz phonon mode, and
a Fano interference effect is clearly observed around $t_{12}\simeq$1
ps, similar to the one observed in the $T$-dependence.\cite{Yusupov1}
Significantly, the spectra appear strongly distorted around $\Delta t_{12}=3.5-4$
ps, showing \textit{asymmetric diagonal }{}``blobs''. After 6 ps
the fluctuations die down and the AM intensity eventually reaches
full amplitude in approximately 60 ps without any significant change
in linewidth and frequency (not shown).

\begin{figure}[h]
\begin{centering}
\includegraphics[width=9cm]{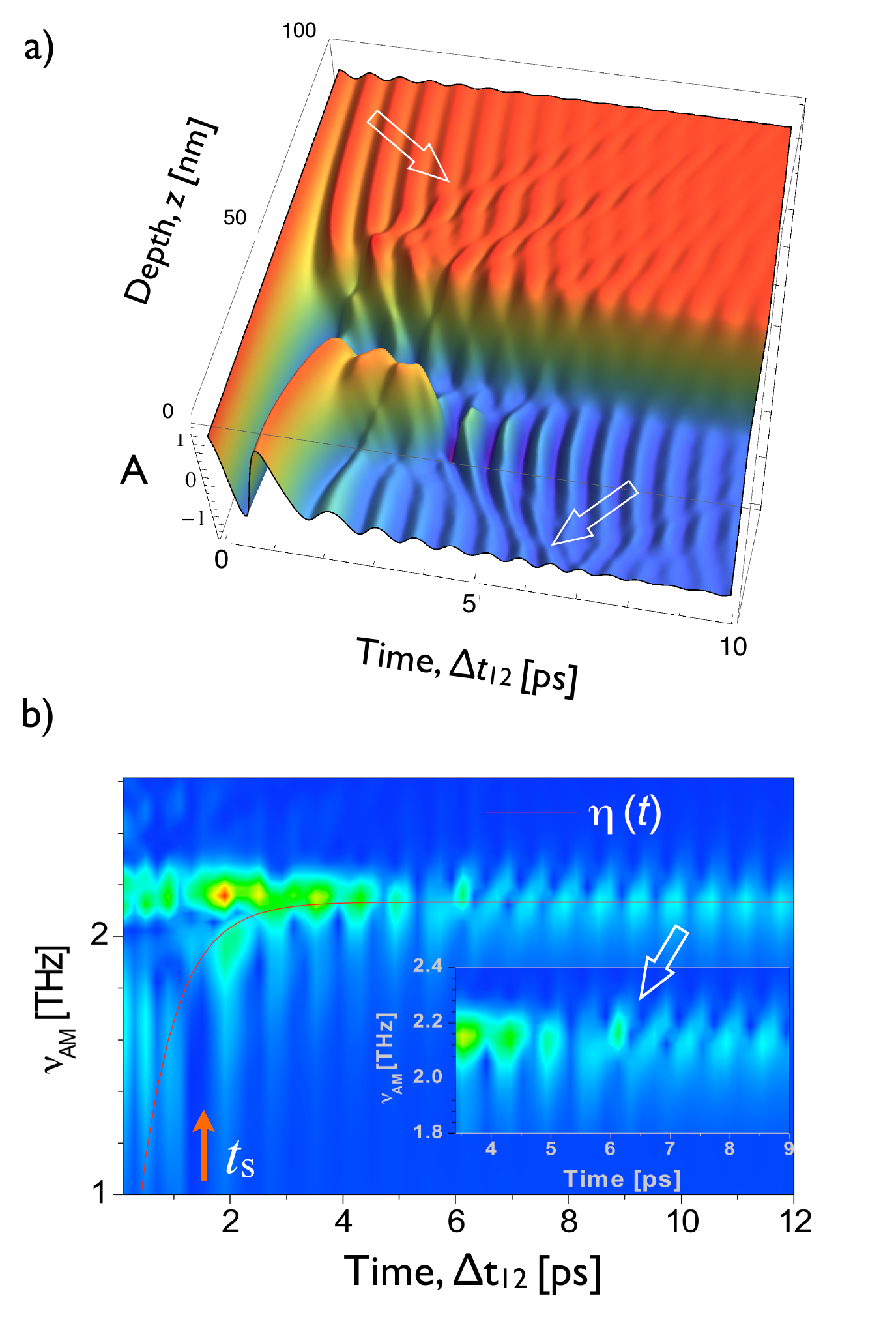} 
\par\end{centering}

\caption{a) The calculated $A(z,t)$ as a function of depth $z$ and $\Delta t_{12}$.
Note the ripples in caused by the annihilation event at $\sim$3.5
ps (arrows). b) The corresponding computed transient reflectivity
$\Delta R(z,t)$ as a function of $\Delta t_{12}$. Note the predicted
diagonal distortion due to the $A(z,t)$-wave reaching the surface,
indicated by the white arrow. The orange arrow points to the critical
slowing down at the critical time of the transition $t_{c}$, i.e.
the bifurcation point. The thin red line shows the function $\eta(t)$.}

\end{figure}

To model the evolution of the system through the SBT, we describe
the field $\Psi$ using a non-linear weakly dispersive Ginzburg-Landau
(GL) model. Neglecting phase fluctuations\cite{phase}, the potential
energy of the system can be described by a double-well, rather than
a \textquotedbl{}Mexican hat\textquotedbl{} potential: \begin{equation}
U=\int dz\left(-\frac{1}{2}(1-\eta)A^{2}+\frac{1}{4}A^{4}+\frac{1}{2}\xi^{2}\left(\frac{\partial A}{\partial z}\right)^{2}\right)\end{equation}
 whose time-dependence is shown schematically in Fig. 3a. Here $A(t,z)=\Delta(t,z)/\Delta_{eq}$
is the time-space dependent amplitude of $\Psi$, normalized to the
equilibrium value $\Delta_{eq}$, and $\xi$ is the coherence length
coupling regions of different $z$. The function $1-\eta(t,z)$ is
a parameter describing the perturbation, akin to the temperature deviation
$(T-T_{c})$ from criticality in usual GL theory. For spatially uniform
$A(t)$, $\eta(t)=\eta(0)\exp(-t/\tau_{A_{QP}})$ as plotted in Fig.
3b). Its exponential form and the parameter $\tau_{A_{QP}}$ are experimentally
determined from fits to Fig. 2b). The time-evolution of $U$ and $\mu=1-\eta$
with $\eta(0)=2$ is shown schematically in Fig. 3a): Before the $D$
pulse, and for large $t$ or $z$, $\eta=0$, so the system resides
in a homogeneously ordered ground state with $|A|=1$. Immediately
after the $D$ pulse, $1-\eta<0$ and the double well potential disappears
in favor of a single energy minimum at $A=0$. %But the system has been prepared in a symmetry broken state with $A\ne 0$, and the viscosity is proven to be low so that the energy is almost conserved. Then $A(t)$ enters the regime of unharmonic, pendulum-like oscillations between positions $\pm A_{max}$ with $A_{max}$ gradually reducing in time from $1$. 
As $1-\eta$ increases and becomes positive, nonzero minima emerge
at $\pm A_{min}=\pm(1-\eta)^{1/2}$, and start to attract the system,
which is soon trapped in one of them, and the symmetry is broken again.
From Eq. {[}1{]}, the equation of motion can then be written as: \begin{equation}
\frac{1}{\omega_{0}^{2}}\frac{\partial^{2}}{\partial t^{2}}A+\frac{\alpha}{\omega_{0}}\frac{\partial}{\partial t}A-(1-\eta)A+A^{3}-\xi^{2}\frac{\partial^{2}}{\partial z^{2}}A=0\end{equation}
 Here $\omega_{0}$ is the angular frequency of the bare ($2k_{F}$)
phonon mode responsible for the CDW formation; the second term describes
its damping $\alpha\le{\Delta\nu_{AM}}/{\nu_{AM}}$. The exponentially
decaying light intensity due to the finite penetration depth of light
is accounted for by the excitation function $\eta'(t,z)=\eta(t)\exp(-z/\lambda)$
where $\lambda$=20 nm is the light absorption depth of TbTe$_{3}$
at 800 nm. Using the experimental values for $\tau_{QP}$, $\nu_{AM}=\tilde{\omega}_{0}/2\pi=2.18$
THz, the linewidth $\Delta\nu_{AM}=0.2$ THz, coherence length $\xi=1.2$
nm \cite{STM} and penetration depth $\lambda=20$ nm, there are no
free parameters and we can compute $A(t,z)$. In Fig. 3b) we first
plot the spatially homogeneous solution with $\xi=0$, with and without
the $P$ pulse. In this particular simulation, the parameters were
chosen to illustrate that a small perturbation of the $P$ pulse causes
the system to revert to a different minimum. With many preceding oscillations,
the final ground state is ergodically uncorrelated with the initial
one, hence the formation of domains is expected under inhomogeneous
conditions. 

The full inhomogeneous solution $A(t,z)$ to Eq. 2 is plotted in Fig.
4a). We see that after $\sim1$ ps, four domains are formed parallel
to the surface with $A(t,z)$ oscillating either around 1 or -1 (orange
or blue respectively), accompanied by the emission of $A(t,z)$-field
waves, which propagate into the sample. At $\sim3$ ps we observe
the fusion of two domain walls, which is accompanied by the emission
of field waves of $A(t,z)$ now propagating towards the surface \emph{and}
into the bulk (arrows). They appear to reach the surface around $\Delta t_{12}\simeq$
4-5 ps which - as we shall see - cause detectable distortions of the
spectra at around 5-6 ps. The appearance of such $A$-waves following
annihilation is qualitatively robust with respect to the parameter
values for TbTe$_{3}$ (see SI). (More $A$-field wave dynamics is
shown in the accompanying movies.)

The calculated reflectivity response detected by the probe $p$ is
given by the difference between the response \emph{with} and \emph{without}
the pump $P$ pulse: $\Delta R(t,\Delta t_{12})\propto\int_{0}^{\infty}[A_{D}^{2}(t,z)-A_{DP}^{2}(t,z,\Delta t_{12})]e^{-z/\lambda}dz$,
where $A_{DP}^{2}(t,z,\Delta t_{12})$ is calculated replacing $\eta\rightarrow\eta_{P}(t)=\eta(0)\exp(-t/\tau_{sp})+\Pi\Theta(t-t_{12})exp[-(t-t_{12})]/\tau_{sp})$.
The exponential term accounts for the probe penetration depth. The
typical value of $\Pi=0.1$, where $\Pi$ is the \emph{P} pulse intensity
relative to the \emph{D} pulse and $\Theta(t-t_{12})$ is a unit step
function (see SI for details). The calculated response for the homogeneous
solution $\Delta R(t,\Delta t_{12})$ is shown by the green curve
in Fig. 3b).

In Fig. 4b) we show the FFT power spectra from $\Delta R(t,\Delta t_{12})$
taking full account of spatial inhomogeneity for D, P and p. The main
features of our data in Fig. 2c) are unmistakably present: oscillations
of $A(t)$ are clearly visible at short times, as well as the hallmark
of the transition itself, namely the critical slowing of the AM oscillations
close to the critical point $t_{c}\simeq1.5$ ps, pinpointing the
exact critical time of the transition $t_{c}$. At this bifurcation
point topological defects are formed. The calculation also reproduces
the softening of the AM for $\Delta t_{12}<2$ ps. After 2ps, the
ripples in $A(t,z)$ discussed above cause a temporal deformation
of the spectral profiles, giving diagonal blobs at $5\sim6$ ps shown
in Fig. 4b). These are remarkably similar to the diagonal spectral
distortions observed in the experimental data in Fig. 2c). A summary
of exhaustive modeling within a wide parameter space is presented
in the SI showing that inhomogeneity without $A$-field waves cannot
cause diagonal deformations, only vertical ones. $ $ The diagonal
distortions in $\omega-t_{12}$ plots are thus unambiguously attributed
to the $A$-field waves created upon the annihilation of defects. 

Identical experiments on three additional microscopically diverse
systems (\emph{2H}-TaSe$_{2}$, K$_{0.3}$MoO$_{3}$ and DyTe$_{3}$)
displaying a 2nd order SBT presented in the SI show that the sequence
of events after the quench: \textit{(i) ultrafast QP gap recovery}
$\rightarrow$ \textit{(ii) $\Psi$-field amplitude fluctuations}
$\rightarrow$ \emph{(iii)} \textit{critical slowing down through}
$t_{s}$ \textit{and (iv) domain creation} $\rightarrow$ \textit{coherent
defect annihilation} is commonly observed in systems which unambiguously
belong to the same universality class (the tellurides and the selenide).
The microscopic properties of the underlying vacuum such as $\lambda$
and $\xi$ change the details. K$_{0.3}$MoO$_{3}$ - which already
shows lack of AM softening in the \emph{T}-induced transition\cite{KMO}-
does not show step (iv), which we attribute to departure from universality.
It is interesting to note that the mechanism described here for topological
defect creation is conceptually and historically related not only
to vortex formation in superconductivity, but also to the Kibble-Zurek
mechanism for the formation of cosmic strings. The $A(t,z)$ -waves
such which we observe after annihilation events have a direct analogue
in the Higgs spontaneous symmetry breaking mechanism. All these models
share a common underlying potential, albeit with different symmetries
of OP and microscopic properties of the underlying vacuum\cite{topology,Varma,Kibble,Higgs}.
A notable distinction of our system is that $\phi$ relaxation is
slow compared to the relaxation of the potential itself, allowing
the collective mode and topological defect dynamics to be clearly
observed. In superconductors, the QP relaxation is beautifully observed\cite{key-1},
but the collective mode is overdamped and unobservable\cite{Gorkov}.
Thus real-time observations and coherent control of the macroscopic
order through the SBT appears to be a rather unique feature of femtosecond
laser experiments on CDW systems.

\textbf{Acknowledgments.} We wish to thank Christoph Gadermaier for
critial reading of the manuscript. Work at Stanford University supported
by the Department of Energy, Office of Basic Energy Sciences under
contract DE-AC02-76SF00515. S.B. acknowledges the support %%%
from the ANR project BLAN07-3-192276.

%
\begin{comment}
\bibitem{Liqcrys}Chuang, I., Durrer, R., Turok N., and Yurke, B.,
Cosmology in the laboratory: Defect dynamics in liquid crystals. \textit{Science}
\textbf{251}, 1336-1342 (1991).

\end{comment}
{}

\begin{thebibliography}{17}
\bibitem{phase} Phase perturbations can be neglected on short timescales,
because they require significant time for nucleation.
\end{thebibliography}
\end{document}